\documentclass[preprint]{iucr}              
     \journalcode{J}              
\usepackage{graphicx} 

\usepackage{pdfpages}
\usepackage{amsmath}
\usepackage{epsfig}
\usepackage{tabularx}
\usepackage{booktabs}
\usepackage{multirow}
\usepackage{textcomp}
\usepackage{makeidx}
\makeindex

\begin{document}                  



\title{Global SAXS data analysis for multilamellar vesicles: Evolution of the scattering density profile (SDP) model}
\shorttitle{Evolving the SDP model}


\author[a]{Peter}{Heftberger}
\author[a]{Benjamin}{Kollmitzer}
\author[b]{Frederick A.}{Heberle}
\author[b,c]{Jianjun}{Pan}
\author[d,e]{Michael}{Rappolt}
\author[d]{Heinz}{Amenitsch}
\author[f]{Norbert}{Ku{\v c}erka}
\author[b,g,h,i]{John}{Katsaras}

\cauthor[a]{Georg}{Pabst}{georg.pabst@uni-graz.at} 

\aff[a]{Instiute of Molecular Biosciences, Biophysics Division, University of Graz, \country{Austria}}
\aff[b]{Biology and Soft Matter Division, Oak Ridge National Laboratory, Oak Ridge, TN, \country{USA}}
\aff[c]{Department of Physics, University of South Florida, Tampa, FL 33620, \country{USA}}
\aff[d]{Institute of Inorganic Chemistry, Graz University of Technology,  \country{Austria}}
\aff[e]{School of Food Science and Nutrition, University of Leeds, \country{United Kingdom}}
\aff[f]{Canadian Neutron Beam Centre, National Research Council, Chalk River, ON, \country{Canada}}
\aff[g]{Joint Institute for Neutron Sciences,  Oak Ridge, TN, \country{USA}}
\aff[h]{Department of Physics and Astronomy, University of Tennessee, Knoxville, TN, \country{USA}}
\aff[i]{Department of Physics, Brock University, St. Catharines, ON, \country{Canada}}









\maketitle                        


\begin{abstract}
We adapted the highly successful scattering density profile (SDP) model, which is used to  jointly analyze  small angle x-ray and neutron scattering data from unilamellar vesicles, for use with data from fully hydrated, liquid crystalline  multilamellar vesicles (MLVs). By using a genetic algorithm, this new method is capable of providing high resolution structural information, as well as determining bilayer elastic bending fluctuations from standalone x-ray data. Important structural parameters, such as bilayer thickness and area per lipid were determined for a series of saturated and unsaturated lipids, as well as binary mixtures with cholesterol. Results are in good agreement with previously reported SDP data, which used both   neutron and x-ray data. The addition of deuterated and non-deuterated MLV neutron data to the analysis improved lipid backbone information, but did not improve, within experimental uncertainties, the structural information regarding bilayer thickness and area per lipid.
\end{abstract}


\section{Introduction}
\label{sec:Introduction}
Phospholipids are a major component of biological membranes, and the structural analysis of pure lipid  membranes is an important area of research, as it can provide valuable insights into membrane function, including how the membrane's mechanical properties can affect lipid\slash protein interactions  \cite{escriba_membranes:_2008,mouritsen_life_2005}. Of the liquid crystalline mesophases formed by phospholipids in aqueous solutions, most effort has been expended to liquid crystalline (fluid) bilayers ($L_{\alpha}$), because of their biological significance. 

Over the years scattering techniques, such as  small angle x-ray and neutron scattering (SAXS, SANS) have been widely used to determine the structural parameters and mechanical properties of biomimetic membranes. With regard to bilayer structure, two important structural parameters are  bilayer thickness and lateral area per lipid $A$ \cite{lee_how_2004,pabst_applications_2010,heberle_model-based_2012}; the latter is directly related to lipid volume and inversely proportional to bilayer thickness. Importantly, $A$  plays a key role in the validation of molecular dynamic (MD) simulations \cite{klauda_simulation-based_2006}, as such its value for different lipids must be accurately known. Over the years for a given lipid, a range of values for $A$ have been reported \cite{kucerka_influence_2007}. Since lipid volumes are determined from independent and highly accurate densitometry measurements \cite{nagle_structure_2000,greenwood_partial_2006,uhrikova_component_2007}, differences in $A$ must therefore result from differences in bilayer thickness. To accurately determine lipid areas, a precise measure of the Luzzati thickness $d_B$ \cite{luzzati_structure_1962}, which is given by the Gibbs dividing surface of the water/bilayer interface \cite{kucerka_lipid_2008}, is needed. Other frequently used definitions of bilayer thickness are the headgroup-to-headgroup thickness $d_{HH}$ and the steric bilayer thickness \cite{pabst_structure_2003}. The latter two bilayer thickness definitions can also be used to determine $A$, however, assumptions regarding headgroup size or distance to the chain/headgroup interface have to be made. 

There are two important issues which one must consider when measuring membrane thickness. Firstly, due to the thermal disorder of fluid bilayers, there is no distinct division between lipid and water, instead a water concentration gradient exists at the membrane's interface. Secondly, x-rays and neutrons are sensitive to different parts of the bilayer. X-rays, for example, are strongly scattered from the electron dense phosphate group which is part of the phosphorylcholine headgroup, hence accurate values for $d_{HH}$ can be obtained. On the other hand, neutrons are scattered by atomic nuclei. Since, hydrogen and its isotope deuterium scatter neutrons with similar efficiency, but 180 degrees out-of-phase with each other (i.e., deuterium's coherent scattering length is positive, and hydrogen's is negative). In the case of protiated lipid bilayers SANS is highly sensitive to locating the hydrogen depleted carbonyl groups.  Importantly, however, is that neutron contrast  can be easily tuned by varying the hydrogen-deuterium content of the water (by varying the $\text{H}_2\text{O}$\slash $\text{D}_2\text{O}$ ratio) or of the bilayer (through the use of deuterated lipids) \cite{pabst_applications_2010}. As mentioned, in the case of protiated lipid bilayers in 100\% $\text{D}_2\text{O}$, neutrons are most sensitive to the lipid's glycerol backbone. Moreover, the Gibbs dividing surface for the apolar\slash polar interface is typically located between the headgroup phosphate and the lipid backbone. Therefore a combined analysis of x-ray and neutron data should yield the most accurate values of $d_B$ and $A$ \cite{kucerka_lipid_2008,kucerka_fluid_2011,pan_molecular_2012}. In this combined data analysis, commonly known as the scattering density profile (SDP) model, the lipid bilayer is represented by volume distributions of quasi-molecular fragments, which are easily converted into electron density or neutron scattering length density distributions by simple scaling (for a given molecular group) the appropriate electron or neutron scattering length density  (see \cite{heberle_model-based_2012}, for a recent review). 

Scattering techniques are also capable of probing membrane elasticity. Lipid bilayers are two-dimensional fluids which exhibit significant bending fluctuations of entropic origin. In multilamellar arrangements, \textit{e.g.}, in MLVs or surface supported multibilayers, this leads to a characteristic power-law decay of the positional correlation function, known as quasi long-range order, with Bragg peaks having characteristic line shapes \cite{liu_diffuse_2004,salditt_thermal_2005,pabst_applications_2010}. Membrane elasticity can, therefore, be determined from  line-shape analysis of the Bragg peaks and the underlying physics of this phenomenon is described by the Caill\'e \cite{caille_remarques_1972} or modified Caill\'e theory (MCT) \cite{zhang_theory_1994}. The resulting fluctuation, or Caill\'e parameter $\eta$ is a function of the bilayer bending modulus and the bulk modulus of interbilayer compression. Due to the higher resolution data in reciprocal space, compared to neutrons, x-rays are better suited for determining the shape of Bragg peaks.
 
Just over a decade ago, Pabst and coworkers were the first to report a full $q$-range analysis of MLV SAXS data using MCT \cite{pabst_structural_2000,pabst_structural_2003}. In that method,  quasi-Bragg peaks and diffuse scattering were both taken into account when analyzing the data, and the electron density profile was modeled by a simple summation of Gaussians representing the electron rich lipid headgroup and electron poor (in relation to the headgroup) hydrocarbon chains. Selected examples of this SAXS method of data analysis can be found in the recent reviews by Pabst and coworkers \cite{pabst_applications_2010,pabst_use_2012}.

The work described here extends the global analysis program (GAP) for MLVs   by making use of SDP's description  of the lipid bilayer. This modified technique, termed herein the SDP-GAP model, has several advantages. Firstly, compared to extruded unilamellar vesicles (ULVs), spontaneously forming MLVs are easier to prepare \cite{heberle_model-based_2012}. Secondly, SDP's description of the bilayer  imparts to GAP the ability to simultaneously analyze SANS and SAXS data, while enabling SDP to determine bending fluctuations, and hence bilayer interactions.

In the present study we also attempted to determine precise values of $d_B$ and $A$ using standalone x-ray data. Such analysis, however, is complicated by the use of more fitting parameters, as compared to GAP, and inherently less scattering contrast, as compared to the SDP model, which makes use of  both SANS and SAXS data. To address these shortcomings we used a genetic algorithm, as an optimization routine, in combination with physical information from other sources in order to reduce the number of parameters needed by the SDP-GAP model. To test the new SDP-GAP model, we analyzed a series of saturated and unsaturated phospholipids, as well as  binary lipid mixtures with cholesterol. Results compare favorably with previously reported data obtained using the SDP model, including the commonly accepted bilayer condensation effect induced by cholesterol. We also include SANS data of protiated  and deuterated palmitoyl-oleoyl phosphatidylcholine (POPC) in our analysis, which gives rise to a better resolved location of the lipid's glycerol backbone. Compared to standalone SAXS analysis, differences in $A$ and $d_B$ values obtained from the SDP-GAP model are observed but the differences are well within experimental uncertainty.

\section{Material and Methods}

\subsection{Sample Preparation} 
\label{subsec:Sample Preparation}
1,2-dipalmitoyl-\textit{sn}-glycero-3-phosphocholine (DPPC),  1-palmitoyl-2-oleoyl-\textit{sn}-glycero-3-phosphocholine (POPC),  1-palmitoyl(d31)-2-oleoyl-\textit{sn}-glycero-3-phosphocholine (POPC-d31),  1-stearoyl-2-oleoyl-\textit{sn}-glycero-3-phosphocholine (SOPC), 1,2-dioleoyl-\textit{sn}-glycero-3-phosphocholine (DOPC) were purchased from Avanti Polar Lipids , Alabaster AL, and cholesterol was obtained from Sigma-Aldrich (Austria). 99.8\% $\text{D}_2\text{O}$ was obtained from Alfa Aesar (Ward Hill, MA). All lipids were used without further purification. 

For x-ray experiments, lipid stock solutions (DPPC, DOPC, SOPC, DOPC) were prepared by dissolving predetermined  amounts of dry lipids in chloroform/methanol (2:1, v/v). Binary mixtures with cholesterol (20\,mol\%) were obtained by mixing lipid stock solutions in the appropriate ratios. Lipid solutions were subsequentially dried under a stream of nitrogen and placed under vacuum for about 12 hours, forming  a thin lipid film on the bottom of glass vials. Films were hydrated using 18\,M$\Omega$/cm water by incubation for 2 hours above the lipid melting temperature, with vortex mixing every 15\,min. The final lipid concentration for each sample was 50\,mg/ml.

For neutron experiments, MLVs  of POPC-d31 at 10\,mg/mL were prepared by weighing 15\,mg of dry lipid powder into 13\,$\times$\,100\,mm glass culture tubes and hydrating with 1.50\,mL $\text{D}_2\text{O}$ preheated to 40\,\textcelsius, followed by vigorous vortexing to disperse the lipid. The resultant MLV suspension was incubated at 40\,\textcelsius\ for 1 hour with intermittent vortexing, and then subjected to 5 freeze/thaw cycles between -80 and 40\,\textcelsius\ to reduce the average number of lamellae and facilitate extrusion \cite{kaasgaard_freeze/thaw_2003,mayer_solute_1985}. A 0.75\,mL aliquot of the MLV sample was then used to prepare unilamellar vesicles (ULVs) using a hand-held miniextruder (Avanti Polar Lipids, Alabaster, AL), assembled with a 50\,nm pore-diameter polycarbonate filter and heated to 40\,\textcelsius. The suspension was passed through the filter 41 times. ULV samples were measured within 24\,h of extrusion. Final sample concentrations were 10\,mg/mL, which allows for sufficient water between vesicles to eliminate the interparticle structure factor, thereby simplifying data analysis.

\subsection{Small angle x-ray scattering}
\label{subsec:Small angle x-ray scattering}
X-ray scattering data were acquired at the Austrian SAXS beamline Elettra Trieste, Italy using 8\,keV photons. Diffraction profiles were detected utilizing a Mar300-image-plate detector (MarResearch GmbH, Norderstedt, Germany) and calibrated using a powder sample of silver behenate. Lipid dispersions were taken up in 1\,mm thick quartz capillaries and inserted into a multi-position sample holder. Lipid dispersions in capillaries were equilibrated for a minimum of 10\,min prior to measurement at a predetermined temperature with an uncertainty of $\pm 0.1$\, \textcelsius\ using a circulating water bath. Exposure time was set to 240\,sec. Scattering patterns were integrated using the program Fit2D \cite{hammersley_fit2d:_1997}. 
Background scattering originating from water and air was subtracted, and data sets were normalized using the transmitted intensity, which was measured by a photodiode placed in the beam stop.

\subsection{Small angle neutron scattering}
\label{subsec:Small angle neutron scattering}
Neutron scattering experiments were performed using the Extended Q-range Small-Angle Neutron Scattering (EQ-SANS, BL-6) beamline at the Spallation Neutron Source (SNS) located at Oak Ridge National Laboratory (ORNL). ULVs were loaded into 2\,mm path-length quartz banjo cells (Hellma USA, Plainview, NY) and mounted in a temperature-controlled cell paddle with an 1\,\textcelsius\ accuracy. In 60\,Hz operation mode, a 4\,m sample-to-detector distance with a $2.5 - 6.1\,\mathring{\text{A}}$ wavelength band was used to obtain the relevant wavevector transfer. Scattered neutrons were collected with a two-dimensional (1\,m\,$\times$\,1\,m) $^3\text{He}$ position-sensitive detector made up of 192\,$\times$\,256 pixels. 2D data were reduced using MantidPlot (http://www.mantidproject.org/). During data reduction, the measured scattering intensity was corrected for detector pixel sensitivity, dark current, sample transmission, and background scattering contribution from the water and empty cell. The one-dimensional scattering intensity $I$ vs. $q$ was obtained by radial averaging of the corrected 2D data.

\subsection{Modelling of phospholipid bilayer} 
\label{subsec:Modelling of phospholipid bilayer}

To analyze the scattering profile of MLVs, we adopted the full $q$-range GAP model of Pabst et al. \cite{pabst_structural_2000,pabst_structural_2003}, which takes into account diffuse scattering originating from positionally uncorrelated
bilayers (scaled by $N_{diff}$):
\begin{equation}
I(q) = \frac{1}{q^2}\left(\left|F(q)\right|^2S(q)(1-N_{diff})+\left|F(q)^2\right|N_{diff}\right),
\label{eq:intensity2}
\end{equation}

where the scattering vector $q = 4\pi \sin \theta / \lambda$,  $\lambda$ is the wavelength, $2\theta$ is the scattering angle relative to the incident beam,  $F(q)$ is the bilayer form factor, and $S(q)$ the inter-bilayer structure factor. For fluid lipid bilayers $S(q)$ is given by the Caill\'e theory, and is described in detail in \cite{caille_remarques_1972,zhang_theory_1994,pabst_structural_2000,pabst_structural_2003}. Averaging over variations in scattering domain size was performed following \cite{fruhwirth_structure_2004}. One of the important parameters determined from fitting $S(q)$ using MCT is the Caill\'e parameter $\eta$, which is a measure of bending fluctuations \cite{pabst_applications_2010}. Instrumental resolution was taken into account by convoluting equation\,(\ref{eq:intensity2}) with the beam profile \cite{pabst_structural_2000,qian_peptide-induced_2011}. Additionally, incoherent background and instrumental artefacts were taken into account by the model.

The form factor is the Fourier transform of the electron density or neutron scattering length density profile. In the present work, we implemented the SDP model \cite{kucerka_lipid_2008} for the bilayer. The basis of the SDP model is the description of the membrane by volume distributions of quasi-molecular fragments. A detailed description of volume probability distribution functions can be found in \cite{kucerka_lipid_2008}. The water-subtracted scattering length density distributions ($\Delta\rho(z)$) are then calculated by scaling the volume probability distributions by the component's total electron densities (for x-rays) or neutron scattering length densities. The  form factor is then calculated as:

\begin{equation}
F(q)=\int{\Delta\rho(z)\exp{[-\mathrm{i}qz]}\,\mathrm{d}z}.
\label{eq:formfactor}
\end{equation}

\begin{figure}
\caption{Illustration of the bilayer parsing scheme (top panel) and volume probability distribution (bottom panel) for DPPC. Data are from experiments carried out in the current study.}
\includegraphics[width=8.8cm]{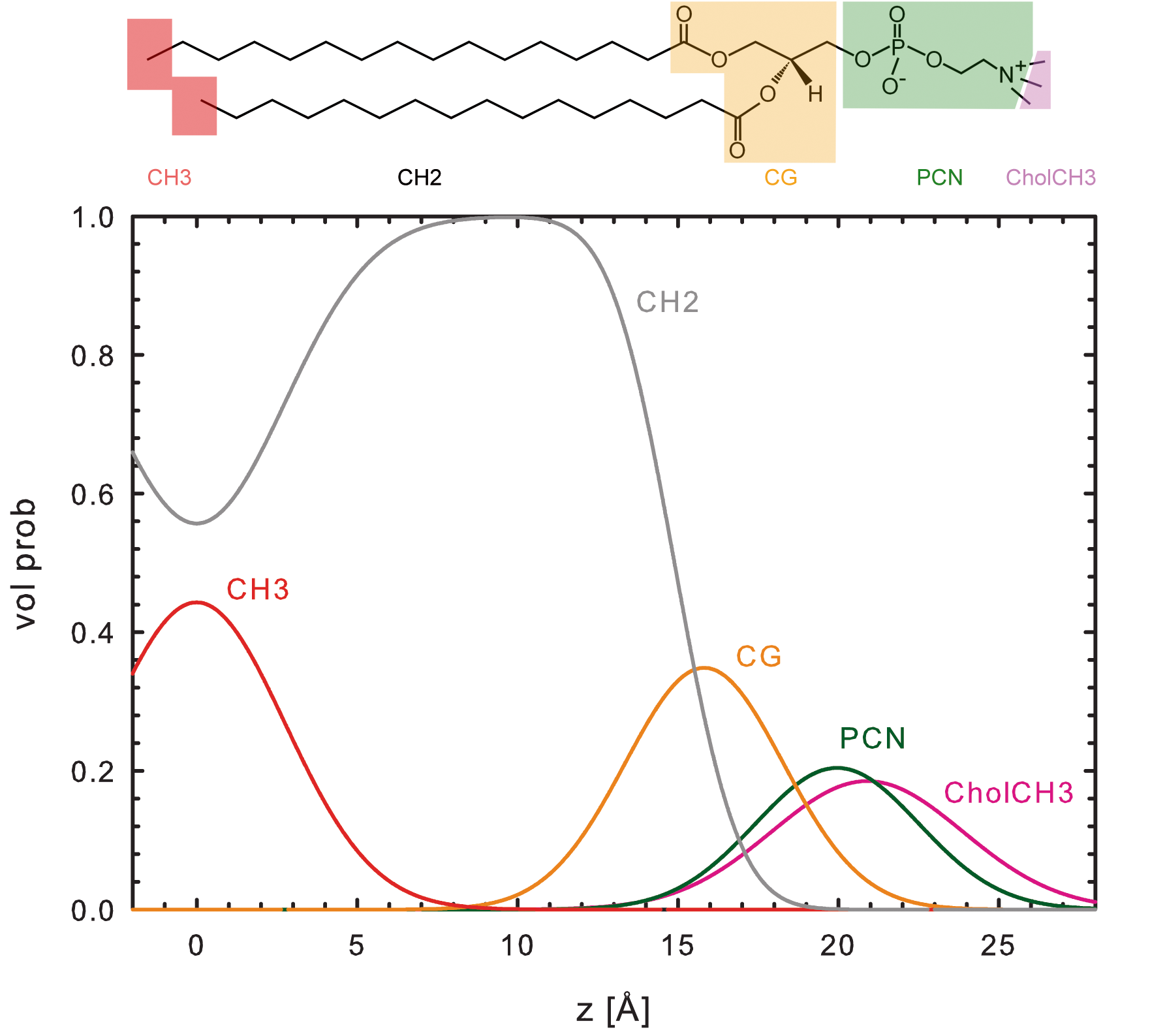}
\label{fig:parsing_scheme}
\end{figure}

Ku\v{c}erka and coworkers originally parsed phosphatidylcholines into the following components: choline methyl (CholCH3); phosphate\,+\,CH$_2$CH$_2$N  (PCN); carbonyl\,+\,glycerol (CG); hydrocarbon methylene (CH2); and hydrocarbon terminal methyl (CH3). An additional methine (CH) group was added for unsaturated hydrocarbon chains. However, the constrast between CH and CH2 is weak, even for SANS \cite{kucerka_lipid_2008}, and effectively zero for SAXS. Hence, our parsing scheme combined the CH with the CH2 group (Fig.\,\ref{fig:parsing_scheme}).

To avoid any non-physical results, the following constraints were adopted from \cite{klauda_simulation-based_2006,kucerka_lipid_2008}. Because of bilayer symmetry, the position of the terminal methyl group $z_{CH3}$ was set to zero and the height of the error function, which describes the hydrocarbon chains, was set to one in order to comply with spatial conservation. The width of the choline methyl group $\sigma_{CholCH3}$ was fixed to $2.98\,\mathring{\text{A}}$, and the width of the error function describing the hydrocarbon chain was constrained within accepted limits ($\sigma_{HC}\in [2.4,2.6]\,\mathring{\text{A}}$) \cite{klauda_simulation-based_2006,kucerka_lipid_2008}.

We also implemented new constraints to aid the standalone x-ray data analysis. Firstly, the distances between the CholCH3 and PCN groups, and the hydrocarbon chain interface ($z_{HC}$) and CG ($z_{CG}$) groups, were not allowed to exceed $2\,\mathring{\text{A}}$. Secondly, volumes of the quasi-molecular fragments, necessary for calculating electron or neutron scattering length densities, were taken from previous reports \cite{kucerka_structure_2005,kucerka_lipid_2008,klauda_simulation-based_2006,kucerka_fluid_2011,greenwood_partial_2006} and allowed to vary by $\pm 20\%$. The total volume of the headgroup components (i.e., CholCH3, PCN, and CG) were constrained to a target value of $331\,\mathring{\text{A}}^3$, as reported in \cite{tristram-nagle_structure_2002}, whereby the value is allowed to deviate from the target value, but in doing-so, incurring a goodness-of-fit penalty.

For lipid mixtures with cholesterol, cholesterol's volume distribution was merged with the CH2 group, following \cite{pan_interactions_2012}. This is justified on the basis of cholesterol's strong hydrophobic tendency, which dictates  its location within the hydrocarbon chain region, and the fact that its hydroxyl group resides in the vicinity of the apolar\slash polar interface \cite{pan_interactions_2012}. In calculating the lipid area for binary mixtures the apparent area per lipid $A=2V_L/d_B$ was used \cite{pan_interactions_2012,pan_effect_2009}.
The bare volume of cholesterol within lipid bilayers was taken to be $630\,\mathring{\text{A}}^3$ \cite{greenwood_partial_2006}. 

\subsection{Determination of structural parameters} 
\label{subsec:Determination of structutral parameters}
Based on volume probability  distributions and scattering length density profiles, membrane structural parameters were defined as follows: (i) the headgroup-to-headgroup distance $d_{HH}$ is the distance between maxima of the total electron density (i.e., the sum of the component distributions);  (ii) the hydrocarbon chain length $d_C$ is the position of the error function representing the hydrocarbon region $z_{HC}$; and (iii) the Luzzati thickness $d_B$ is calculated from the integrated water probability distribution \cite{kucerka_lipid_2008}:
\begin{equation}
d_B = d -2\int_0^{d/2}P_W(z)\,\mathrm{d}z.
\label{eq:luzzati_thickness}
\end{equation}

The volume distribution function of water was previously defined in \cite{kucerka_lipid_2008}
\begin{equation}
P_W(z)=1-\sum P_i(z),
\label{eq:water_model}
\end{equation}
where i indexes the lipid component groups (i.e., CholCH3, PCN, CG, CH2, and CH3). In order to increase the robustness of the analysis for $d_B$, $P_W$ obtained from the SDP analysis was fitted with an error function, thus giving greater weight to the region close to the lipid headgroup (due to the higher x-ray contrast), as compared to the hydrocarbon chain region. We also attempted to include the $P_W$ model function in the SDP fit, however results were not satisfactory. The area per lipid is then given by \cite{kucerka_lipid_2008}:
\begin{equation}
A = \frac{2V_L}{d_B},
\label{eq:area_lipid}
\end{equation}
where $V_L$ is the molecular lipid volume determined by separate experiments. Finally, the thickness of the water layer was defined as
\begin{equation}
d_W = d-d_B.
\label{eq:thickness_water}
\end{equation}
Unless otherwise stated, experimental uncertainties of all structural parameters, including literature values, are $\pm 2\%$.

\subsection{Fitting Procedure} 
\label{subsec:Data Analysis}
Due to the large number of adjustable parameters (i.e., 21) and our goal to apply the SDP-GAP model  to standalone x-ray data, we chose to use a genetic algorithm in the optimization routine. The main benefit of this algorithm, compared to simple gradient descent routines or more sophisticated optimization algorithms (e.g., Levenberg-Marquardt), is that the fitting procedure does not easily fall into  local minima \cite{goldberg_genetic_1989}. Several hundred generations with populations of ~ 2000 individuals were tested for their fitness, defined here as the reduced chi-squared ($\chi^2$) value, which is equal to the sum of the squared residuals divided by the degrees of freedom \cite{press_numerical_2007}. If $\chi^2$ does not change after 100 generations, the optimization is assumed to have converged and the procedure is terminated. Application of genetic algorithms comes with a greater computational cost, but are efficiently  using parallel processing techniques. For the present study, all routines were encoded in IDL (Interactive Data Language), using the SOLBER optimization routine \cite{rajpaul_genetic_2012}. Typical runtimes for one x-ray scattering profile were between three and five hours on a six core machine (Intel Xeon 2.67\,GHz).

\section{Results and Discussion}

\subsection{X-ray standalone data}

\begin{figure}
\caption{SDP-GAP analysis of SOPC MLVs at 30\,\textcelsius. Panel A compares the SDP-GAP  (black line) and GAP models  (red dashed line) to experimental data (grey circles). The inset to the figure compares the corresponding electron density profiles. Panel B shows the volume probability distribution (left hand side) and the electron density distributions of the defined quasi-molecular fragments (right hand side).}
\includegraphics[width=8.8cm]{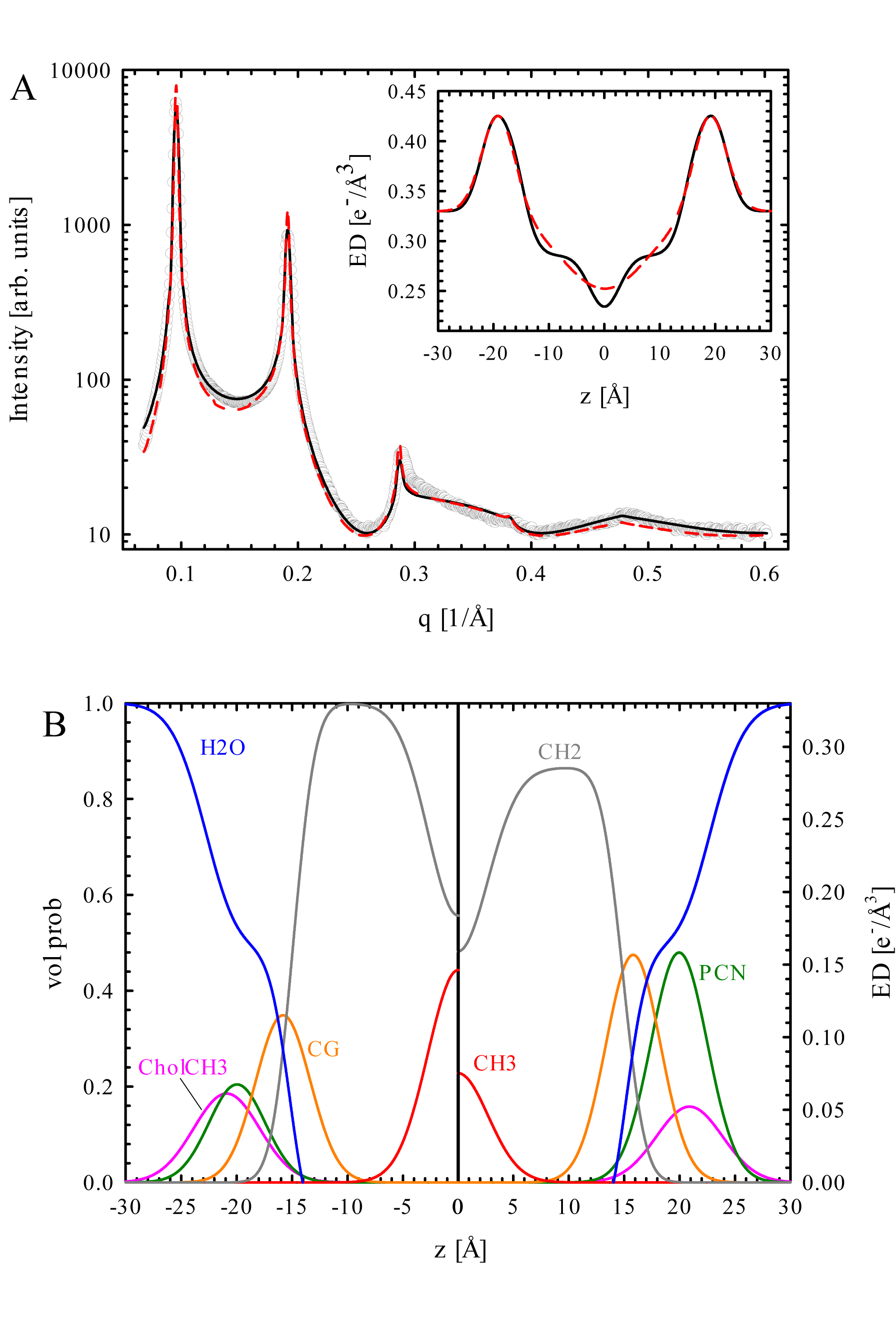}
\label{fig:SOPC-fit}
\end{figure}

The SDP-GAP model was tested on SAXS data obtained from single component $L_{\alpha}$ lipid bilayers and selected binary mixtures of phosphatidylcholines with cholesterol. As an example of our analysis, we present results for SOPC bilayers with five lamellar diffraction orders (Fig.\,\ref{fig:SOPC-fit}). Fits from all other bilayers, including tables with structural parameters, are given in the supplemental material (Figs.\,S1--S3, Tab.\,S1). All SAXS patterns showed significant diffuse scattering, originating from membrane fluctuations common to $L_{\alpha}$ bilayers. In particular, bending fluctuations lead to a rapid decrease in diffraction peak amplitudes as a function of  $q$, and $quasi$-Bragg peaks with characteristic line-shapes. Such effects are accounted for in the structure factor used. We found good agreement between the SDP-GAP model and experimental SOPC data ($\chi^2=0.78$). Fits from other MLV systems yielded similar $\chi^2$ values (Tab.\,S1). Omitting the constraints introduced in section \ref{subsec:Modelling of phospholipid bilayer} led to slightly improved $\chi^2$ values, but produced non-physical  results.

Results from the SDP-GAP model were compared to those from the GAP model. GAP data was in reasonable agreement with the experimental data (Fig.\,\ref{fig:SOPC-fit}), albeit with poorer fit statistics ($\chi^2 = 4.78$), which could be attributed to the small deviations between the various Bragg peaks. Despite the good fits produced using the GAP model, the structural features obtained from SDP-GAP analysis are significantly richer (Fig.\,\ref{fig:SOPC-fit}, lower panel). This point is illustrated by the total electron density shown in the inset to Fig.\,\ref{fig:SOPC-fit}, where the methyl trough is smeared out in the GAP electron density profile.

Table\,\ref{tab:structural_param} provides the main structural parameters obtained from  SDP-GAP and GAP analysis of the same data, as well as literature values obtained from  SDP analysis (i.e., joint refinement of SAXS and SANS data). Calculation of structural parameters using the GAP model is detailed in  \cite{pabst_structure_2003}.  Our results using the SDP-GAP model are in good quantitative agreement with the reference data. Deviations with the GAP model are, however, larger (though still reasonable) due to the  simplified  electron density model that was used. Interestingly, in the case of some lipids, we also find significant differences for the fluctuation parameter. They are the result of the form factor, which modulates peak decay. It therefore stands to reason that the  better fits of the experimental data by the SDP-GAP model should result in more accurate $\eta$ values.

\begin{figure}
\caption{Comparing SDP-GAP and GAP fits to data from SOPC MLVs, with 20\,mol\% cholesterol at 30\,\textcelsius. Nomenclature is same as in Fig.\,\ref{fig:SOPC-fit}.}
\includegraphics[width=8.8cm]{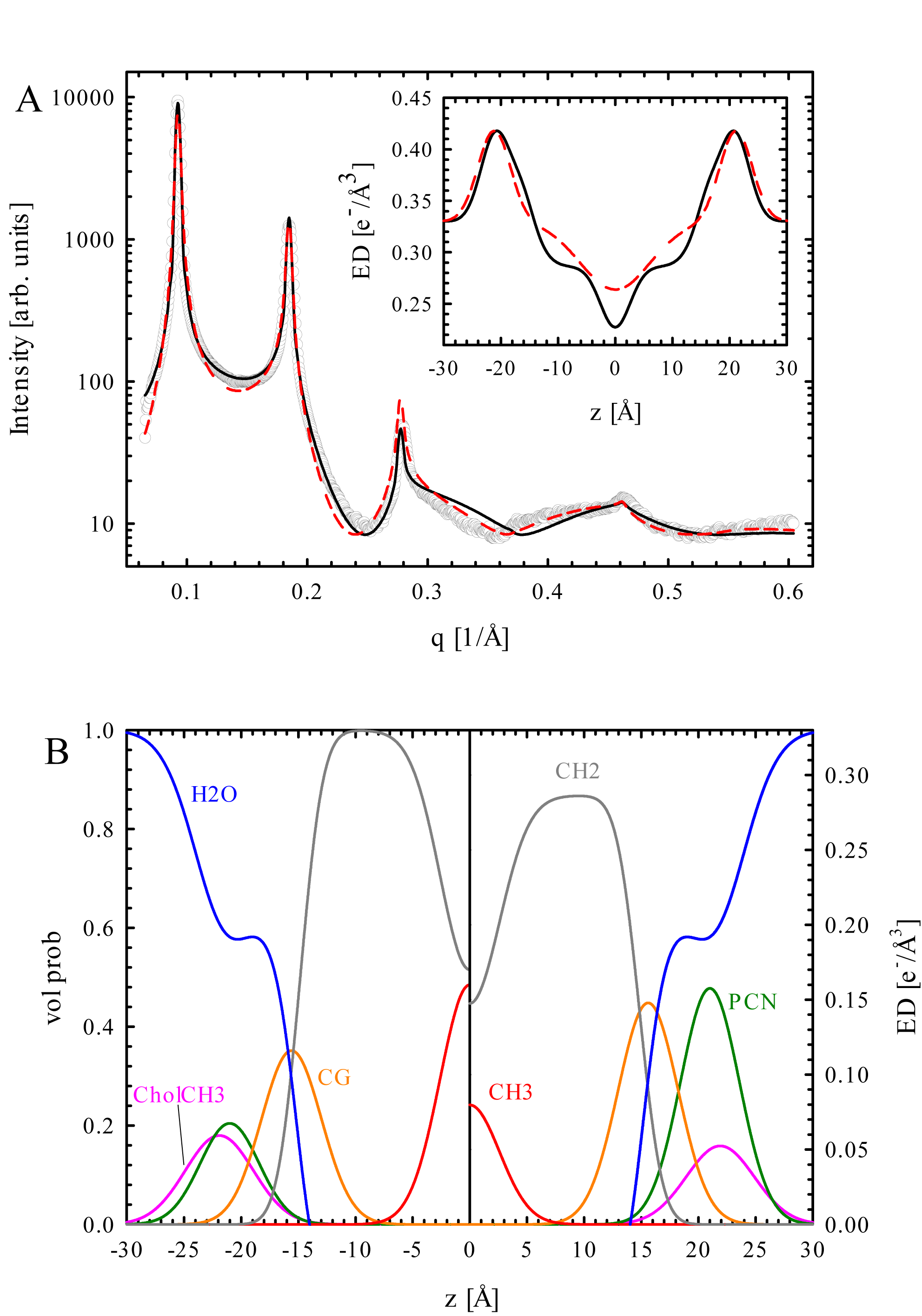}
\label{fig:SOPCuCHOL-fit}
\end{figure}

We further tested the SDP-GAP model using the same lipid systems, but this time with the addition  of 20\,mol\% cholesterol. Cholesterol is abundant in mammalian plasma membranes and is well known for the condensing effect it has on lipid bilayers, which at the molecular level is explained by the  umbrella model \cite{huang_microscopic_1999}. In scattering studies, this effect shows up as an increase in $d_B$ and a concomitant decrease in $A$, as well as reduced  bending fluctuations (see, e.g. \cite{hodzic_differential_2008}). Fig.\,\ref{fig:SOPCuCHOL-fit} shows the fits to data from SOPC\slash cholesterol membranes. The SDP-GAP model is able to describe the more pronounced higher diffraction orders resulting from cholesterol's  presence. Our results show that cholesterol shifts the PCN and CholCH3 groups further away from the bilayer center (Fig.\,\ref{fig:SOPCuCHOL-fit} bottom panel, Tab.\,\ref{tab:effect_cholesterol}, Tab.\,S2), in good agreement with previous reports  \cite{pan_interactions_2012}. On the other hand, we could not observe a significant shift of the CG group from the bilayer center, nor a higher value for hydrocarbon chain thickness (Tab.\,\ref{tab:effect_cholesterol}, S2).

Structural parameters for all lipid mixtures are reported in Tab.\,\ref{tab:effect_cholesterol}. In agreement with previous reports, the addition of cholesterol causes  $A$ to decrease, and $d_B$ and $d_{HH}$ to increase \cite{hung_condensing_2007,kucerka_effect_2008,pan_cholesterol_2008,hodzic_differential_2008,pan_interactions_2012}. Compared to other membrane systems, bending fluctuations in DPPC bilayers experience a greater degree of damping when cholesterol is induced, in agreement with the notion that cholesterol preferentially associates with saturated hydrocarbon chains \cite{pan_effect_2009,pan_cholesterol_2008,ohvo-rekila_cholesterol_2002}. This effect is smaller for lipids having one monounsaturated chain (i.e., SOPC and POPC), and is completely absent when a second monounsaturated chain is introduced (e.g.\,DOPC). This  latter finding is in good agreement  with studies which found  no change in the bending rigidity of DOPC bilayers in the absence or presence of cholesterol \cite{pan_cholesterol_2008}.

SOPC\slash cholesterol mixtures were also analyzed with the GAP model. Although reasonable fits are obtained (Fig.\,\ref{fig:SOPCuCHOL-fit}, $\chi^2_{\textrm{SDP-GAP}}=1.04$, $\chi^2_{\textrm{GAP}}=3.93$) differences in structural parameters when comparing  GAP data  with SDP-GAP data are more pronounced. For example, the total electron density profiles  shows clear deviations in the acyl chain and headgroup regions. Cholesterol increases the asymmetry of the electron density distribution in the headgroup region, as determined from the SDP-GAP model, an effect that is not captured by the single headgroup Gaussian of the GAP model. As a result, parameters such as area per lipid ($A_{\textrm{SDP-GAP}}=60.7\,\mathring{\text{A}}^2$, $A_{\textrm{GAP}}=57.4\,\mathring{\text{A}}^2$) and hydrocarbon chain length ($d_{C,\textrm{SDP-GAP}}=14.9\,\mathring{\text{A}}$, $d_{C,\textrm{GAP}}=17\,\mathring{\text{A}}$) are different between the two two methods, whereas values for head-to-headgroup thickness ($d_{HH,\textrm{SDP-GAP}}=42.1\,\mathring{\text{A}}$, $d_{HH,\textrm{GAP}}=42.3\,\mathring{\text{A}}$) and the Caill\'e parameter ($\eta_{\textrm{SDP-GAP}}=0.05$, $\eta_{GAP}=0.04$) are in reasonably good agreement.

\subsection{Addition of SANS data}
SANS data were obtained from  POPC and POPC-d31 MLVs and ULVs in pure  $\text{D}_2\text{O}$ to see whether or not additional information substantial alters the  results. The protocol devised by Ku\v{c}erka and coworkers used SANS data from proteated bilayers at  different  $\text{H}_2\text{O}$/ $\text{D}_2\text{O}$ contrasts \cite{kucerka_lipid_2008}.

\begin{figure}
\caption{Results of simultaneous SAXS and SANS data analysis of data from  POPC ULVs and MLVs at 30\,\textcelsius. Panel A shows SANS data of POPC (circles) and POPC-d31 (triangles) MLVs, and panel B the corresponding data obtained from ULVs (same symbols). Solid lines are best fits to the data using the SDP-GAP model.  The inserts in panel A and B show the corresponding SAXS fits and neutron length density profiles for POPC (left) and POPC-d31 (right), respectively. Panel C shows the changes in volume distributions from  SAXS-only analysis (dashed black lines, same nomenclature as in Figs.\,\ref{fig:SOPC-fit} and \ref{fig:SOPCuCHOL-fit}) to a simultaneous SAXS\slash SANS analysis (colored lines).}
\includegraphics[width=8.8cm]{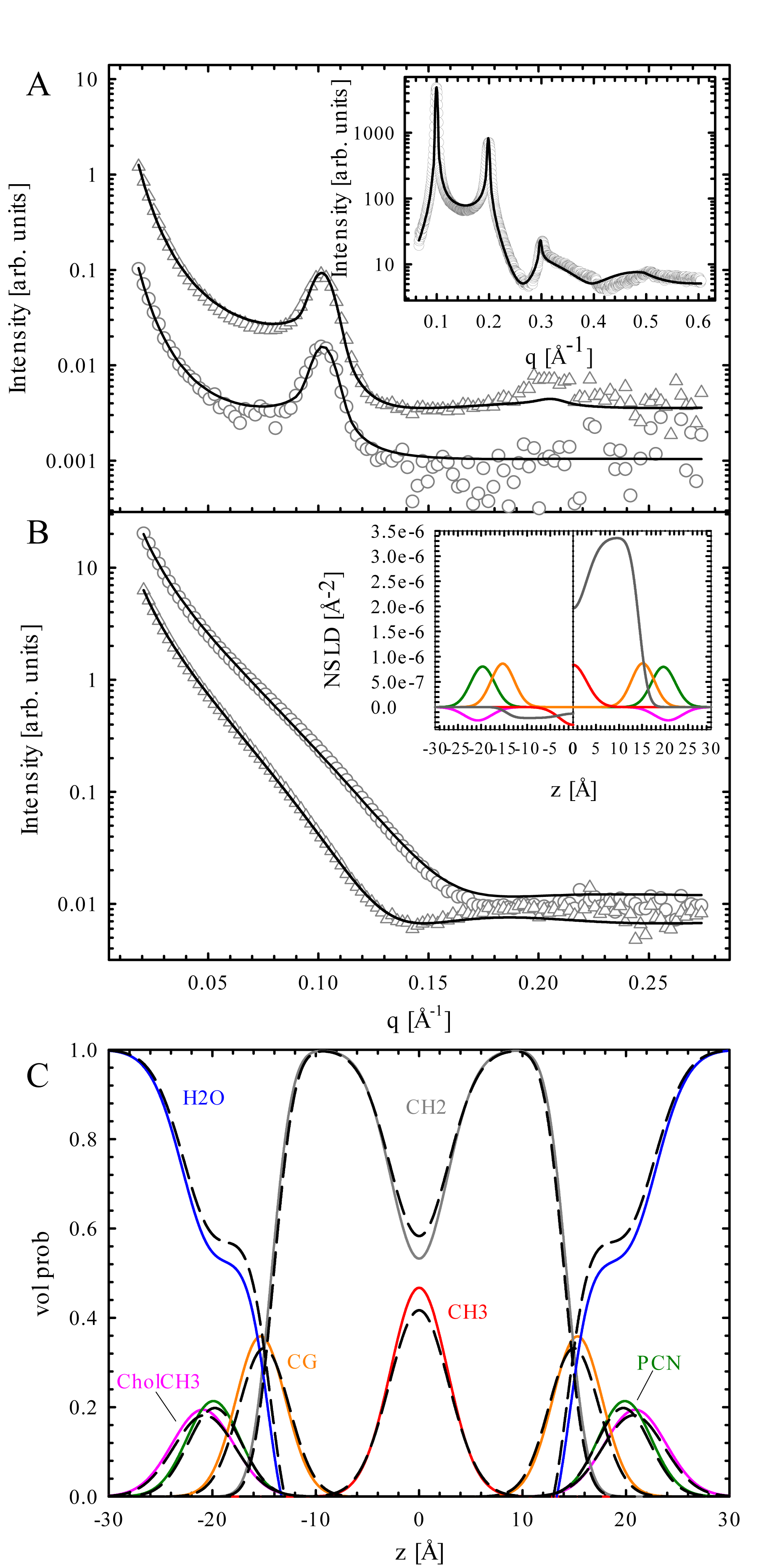}
\label{fig:POPC-fit}
\end{figure}

Replacing H with D shifts the neutron scattering length density (NSLD) profile of the hydrocarbon region from negative to positive values (Fig.\,\ref{fig:POPC-fit}B, insert). Hence, relative to  $\text{D}_2\text{O}$ with SLD $= 6.4 \cdot 10^{-14}\, \text{cm}/\mathring{\text{A}}^3$, the  hydrocarbon chain region contrast is significantly altered. This change in contrast manifested itself by producing two  additional Bragg peaks in the case of  POPC-d31 MLVs, compared to their proteated counterparts (Fig.\,\ref{fig:POPC-fit}A). Similarly, ULV data show a shift of the minimum at low $q$  to higher $q$ vectors for POPC compared to POPC-d31 (Fig.\,\ref{fig:POPC-fit}B), which is also attributed to the change in contrast of the deuterated lipids in $\text{D}_2\text{O}$.

We used SDP-GAP to simultaneously analyze SAXS data in several combinations with SANS data: (i) protiated MLVs; (ii) deuterated MLVs; and (iii) all four SANS data sets (i.e., deuterated and protiated  MLVs and ULVs). We also fit  all MLV data sets simultaneously, and all ULV data sets separately. In doing so, results differed slightly from case (iii), except that case (iii) yielded a lower error for the different  structural parameters. Fit results are shown in Fig.\,\ref{fig:POPC-fit} and the determined structural parameters are summarized in Tab.\,\ref{tab:simultaneous_analysis} and Tab.\,S3. The addition of a single SANS data set produced significant variations in the structural parameters, causing them to deviate from values determined from standalone SAXS analysis and those from literature. This disagreement was rectified by either including  both MLV data sets, or all MLV and ULV data sets in the analysis. In the latter case, significant differences, compared to the standalone SAXS analysis, are found regarding the positions of the  CG group $z_{ZG}$ and $d_C$. This can be understood in terms of the  better neutron contrast of the lipid backbone. Changes in volume distribution functions are shown in Fig.\,\ref{fig:POPC-fit}C. The changes to $A$ and $d_B$ are within the measurement error, and consequently insignificant. We thus conclude that the addition of SANS data helps to improve the location of  the CG group and $d_C$, but offers negligible improvements to values of $A$ and $d_B$.

\subsection{Conclusion}
We have modified the full $q$-range SAXS data analysis, which used a simplified  electron density profile \cite{pabst_structural_2000}, by replacing it with a high resolution representation of scattering density profiles, which are based on volume distributions of quasi-molecular fragments \cite{kucerka_lipid_2008}. The new SDP-GAP is a hybrid model of GAP and SDP that combines the advantages offered by each model into one.  The SDP-GAP model can be applied to MLV and ULV data, and is capable of simultaneously analyzing SAXS and SANS data. An advantage of this new hybrid model is that MLVs are spontaneously formed membrane systems, and their analysis opens up new opportunities for the study of bilayer interactions and membrane mechanical properties (e.g., elasticity) \cite{pabst_applications_2010}.

An additional feature of this new model is its ability  to obtain high resolution structural information from standalone SAXS data. This is achieved by implementing an optimization routine based on a genetic algorithm, which is able to deal with the large number of adjustable parameters needed, even though additional constraints and input parameters were applied in order to limit parameter space. Compared to the GAP and SDP models, which use Levenberg-Marquardt and downhill simplex optimization routines, respectively, the computational effort required by the SDP-GAP model is significantly higher. Typical CPU times on parallel processors are on the order of a few hours, as compared to a few minutes for SDP or GAP. However, an advantage is that the genetic algorithm prevents the optimization routine from stalling in local minima.  By using different seeds for the random number generator, for a given data set, robust results with good convergence are readily obtained

We then tested the SDP-GAP model using different saturated and unsaturated phosphatidylcholine bilayers, with and without  cholesterol. Results for $d_B$ and $A$ are in good agreement with previous reports using the SDP model, although we note that the position and width of the CG groups are subject to greater variabilities due to the lower x-ray contrast of this particular group. This inadequacy was, however, rectified by including ULV SANS data. MLV SAXS data combined with ULV SANS data of POPC and  POPC-d31 bilayers resulted in improved results for both the position of the  CG group and  hydrocarbon chain thickness.  (Fig.\,\ref{tab:simultaneous_analysis}C, Tab.\,\ref{tab:simultaneous_analysis}). However, the values of $A$ and $d_B$ remained practically unchanged.






\ack{This work was supported by the Austrian Science Fund FWF, Project No. P24459-B20 (to G.P.). Support was received from the Laboratory Directed Research and Development Program of Oak Ridge National Laboratory (to J.K.), managed by UT-Battelle, LLC, for the U.S. Department of Energy (DOE). This work acknowledges additional support from the Scientific User Facilities Division of the DOE Office of Basic Energy Sciences, for the EQ-SANS instrument at the ORNL Spallation Neutron Source. This facility is managed for DOE by UT-Battelle, LLC under contract no. DE-AC05-00OR2275.}






\renewcommand*{\thefootnote}{\fnsymbol{footnote}}

\begin{table}
\caption{Comparison of structural parameters.}
\begin{tabularx}{\textwidth}{XXXXX}
   &   &   SDP-GAP &  GAP  & SDP\footnotemark[1] \\
\hline
\multirow{5}{*}{DPPC (50\,\textcelsius)}  &   $A[\mathring{\text{A}}^2]$   &   63.1   &  61.8  &  63.1 \\
  &  $d_B[\mathring{\text{A}}]$         &   39.0 &  n.a.  &  38.9 \\
	&  $d_{HH}[\mathring{\text{A}}]$    &   37.9  &  37.3  &  38.4 \\
	&  $d_C[\mathring{\text{A}}]$   &   13.9 &  14.5  &  14.2 \\
	&  $\eta$    &   0.08  &  0.067  &  n.a. \\
\hline
\multirow{5}{*}{POPC (30\,\textcelsius)}  &   $A[\mathring{\text{A}}^2]$   &   65.4  &  64.3  &  64.4 \\
  &  $d_B[\mathring{\text{A}}]$         &   38.4 &  n.a.  &  39.0 \\
	&  $d_{HH}[\mathring{\text{A}}]$    &   37.3 &  37.0  &  36.5 \\
	&  $d_C[\mathring{\text{A}}]$   &   14.0  &  14.4  &  14.4 \\
	&  $\eta$    &   0.06  &  0.056  &  n.a. \\
	\hline
\multirow{5}{*}{SOPC (30\,\textcelsius)}  &   $A[\mathring{\text{A}}^2]$   &   66.3  &  60.3  &  65.5 \\
  &  $d_B[\mathring{\text{A}}]$         &   39.5  & n.a.  &  40.0 \\
	&  $d_{HH}[\mathring{\text{A}}]$    &   38.7  &  40.7  &  38.6 \\
	&  $d_C[\mathring{\text{A}}]$   &   14.6 &  16.2  &  15.0 \\
	&  $\eta$    &   0.06  &  0.08  &  n.a. \\
	\hline
\multirow{5}{*}{DOPC (30\,\textcelsius)}  &   $A[\mathring{\text{A}}^2]$   &   67.6  &  69.7  &  67.4 \\
  &  $d_B[\mathring{\text{A}}]$         &   38.5  &  n.a.  &  38.7 \\
	&  $d_{HH}[\mathring{\text{A}}]$    &   36.9  &  36.1  &  36.7 \\
	&  $d_C[\mathring{\text{A}}]$   &   14.2   &  13.9  &  14.4 \\
	&  $\eta$    &   0.1  &  0.1  &  n.a. \\
\end{tabularx}
\label{tab:structural_param}
\footnotetext[1]{from\cite{kucerka_lipid_2008,kucerka_fluid_2011}.}
\end{table}

\begin{table}
\caption{Structural parameters from the SDP-GAP model  of lipid bilayers containing 20\,mol\% cholesterol.}
\begin{tabularx}{\textwidth}{lXXXXX}      
Lipid    & $A[\mathring{\text{A}}^2]$  & $d_B[\mathring{\text{A}}]$ & $d_{HH}[\mathring{\text{A}}]$  & $d_C[\mathring{\text{A}}]$ &  $\eta$  \\
\hline
DPPC(50\,\textcelsius)   & 61.2   & 40.1    & 42.3  & 14.2     & 0.02    \\
POPC(30\,\textcelsius) & 63.1    & 39.8    & 40.3   & 14.3      & 0.05   \\
SOPC(30\,\textcelsius)  & 60.6    & 40.5   & 42.1(42.1)\footnotemark[2]   & 14.9(16.1)\footnotemark[2]     & 0.05    \\
DOPC(30\,\textcelsius) & 66.2    & 39.4     & 40.9(39.0)\footnotemark[2]   & 13.5(14.6)\footnotemark[2]      & 0.14    \\
\end{tabularx}
\label{tab:effect_cholesterol}
\footnotetext[2]{from \cite{pan_effect_2009}.}
\end{table}

\begin{table}
\caption{Structural parameters for POPC using different  combinations of SAXS and SANS data.}
\begin{tabularx}{\textwidth}{XXllXX}
     &  SAXS\footnotemark[1]  & n-MLV$_\text{u}$\footnotemark[2]  & n-MLV$_\text{d}$\footnotemark[3] & all data\footnotemark[4]  &  SDP\footnotemark[5] \\
\hline
$A[\mathring{\text{A}}^2] $   & 65.4   &  64.9    &  63.1  & 63.6  &  64.4   \\
$d_B[\mathring{\text{A}}]$  & 38.4     &  38.7    &  39.8  & 39.5  &  39.0  \\
$d_{HH}[\mathring{\text{A}}]$  & 37.3  &  37.1    &  37.3  & 37.5  &  36.5  \\
$d_C[\mathring{\text{A}}]$  &  14.0    &  14.6    &  14.4  & 14.3  &  14.4 \\
$z_{CG}[\mathring{\text{A}}]$  & 15.0  &  15.3    &  15.4  & 15.3  &  15.3  \\
\end{tabularx}
\label{tab:simultaneous_analysis}
\footnotetext[1]{Results obtained using SAXS data only.}
\footnotetext[2]{SAXS (POPC-MLV) and SANS (POPC-MLV) data.}
\footnotetext[3]{SAXS (POPC-MLV) and SANS (POPC-d31-MLV) data.}
\footnotetext[4]{SAXS (POPC-MLVs) and SANS (POPC-ULVs\slash MLVs, POPC-d31-ULVs\slash MLVs) data.}
\footnotetext[5]{from \cite{kucerka_fluid_2011}}
\end{table}

\end{document}